\documentclass[12pt]{revtex4}
\usepackage{amsmath,amssymb,bm}
\usepackage{graphicx}
\usepackage{hyperref,xcolor}

\DeclareGraphicsExtensions{.pdf,.png,.jpg,.eps}

\begin{document}

\title{Weyl's theorem and the theoretical origin of Boson peak in two-dimensional scalar spring network}
\author{Cunyuan Jiang}
\email{cunyuanjiang@zzu.edu.cn}
\affiliation{School of Physics, Zhengzhou University, Zhengzhou 450001, Henan, China}

\begin{abstract}
The origin of the anomalous low-frequency vibrational density of states
(VDOS) universally observed in amorphous solids has been debated for
decades. In this work, we model amorphous solids as a zero-temperature
scalar spring network, the simplest model that captures both the
amorphous structure and the features of the anomalous VDOS. Using this
scalar network, we mathematically demonstrate that disorder can lead to a
shift of the VDOS toward low frequencies, thereby accumulating the
anomalous part there. Disorder is defined through the topological
connectivity of the network. The validity and universality of the
theoretical framework are general and valuable for real-world scenarios
because it rests only on Weyl's theorem for Hermitian matrices and is
free of perturbation and phenomenological assumptions. This work strips
away complex factors and reveals, at the level of mathematical
principles, a profound and clean mechanism for the origin of the
anomalous VDOS in amorphous solids.
\end{abstract}

\maketitle

\section{Introduction}
Vibration is a fundamental dynamical behavior of materials, governing
their acoustic, thermal, and functional properties. Our understanding of
vibrations in solids is rooted in Debye's theory\cite{debye}, which
establishes the low-frequency scaling of the vibrational density of
states (VDOS) for a solid of dimensionality \(w\):
\(\rho(\omega)\propto\omega^{w-1}\). In Debye's framework, discrete
translational symmetry, as in crystals, is essential because it defines
the wave vector and enables analysis based on it\cite{Kittel2004}.
Amorphous solids lack this discrete translational symmetry. Since the
seminal work by Zeller \textit{et al.} in the 1970s\cite{PhysRevB.4.2029},
the anomalous low-frequency VDOS that universally violates Debye's
scaling law in amorphous solids, usually referred to as the Boson peak
(BP), has attracted broad attention.

Over the past decades, studies ranging from ordinary
glasses\cite{Ding2025,PhysRevLett.99.035503,PhysRevB.77.214309,T2023}
to granular glasses\cite{PhysRevB.98.174207,Zhang2017,PhysRevResearch.3.L032067}
and molecular glasses\cite{Gj2023}, and interpretations ranging from a
softened Van Hove peak\cite{PhysRevLett.106.225501} to dynamical
objects\cite{Hu2022,PhysRevLett.133.188302,C2SM26789F,C2SM27533C,Corrado2020,PhysRevLett.108.095501,PhysRevResearch.6.023053}
and recent string resonances\cite{Xi2026,bingyu2026,Jiang_2024}, have
provided a relatively clear picture of the properties of the BP.
Despite extensive and detailed observations, the origin of the BP in
amorphous solids, especially from the theoretical side, remains highly
controversial, as commented in Ref.\cite{Xi2026}. The controversy stems
largely from the absence of a general theoretical framework that
explains the origin of the BP from a fundamental principle.

The theory of inhomogeneous elasticity is perhaps the most widely
disseminated framework for interpreting the origin of the
BP\cite{PhysRevLett.100.137402,PhysRevLett.98.025501}, although it has
been questioned for underestimating microscopic details due to its
perturbative approach\cite{PhysRevLett.123.055501,PhysRevLett.130.236101}.
Moreover, the observation of a BP in hyperuniform glasses\cite{Xi2026},
where elasticity is homogeneous, strongly indicates that inhomogeneous
elasticity is not even necessary for producing a BP. In addition to
perturbatively modeling disorder as inhomogeneous elasticity, another
class of approaches models experimentally and simulatively observed
dynamical objects phenomenologically within field-theoretical
frameworks\cite{Jiang_2024,PhysRevB.91.094102,bingyu2026,jiang20242,p95w-9wjc}.
These models with phenomenological assumptions are successful to a
certain extent compared with experiments and simulations, but a general
theory without phenomenological input for revealing the origin of the BP
is still desirable.

Amorphous solids are generally complex systems involving various
factors that affect their dynamical properties, such as
temperature\cite{PhysRevLett.104.067402,doi:10.1073/pnas.252786999},
slow structural relaxation\cite{aapnas2012,Zhao2025}, and
inhomogeneity\cite{Jiang_2026,qx17-q53j}. A general theoretical model
must grasp the most dominant and fundamental feature of amorphous solids
that leads to the BP. The dominant features are the deepest
distinguishing marks of various theories. The inhomogeneous elasticity
approach\cite{PhysRevLett.100.137402,PhysRevLett.98.025501,Jiang_2026,qx17-q53j}
captures mesoscopic spatial inhomogeneity but loses particle-level
microscopic details. At the other limit, pure particle-level dynamical
simulations provide full microscopic details at the price of sacrificing
analyticity and theoretical readability. Balancing feature salience and
detail accuracy, the concept of local inversion symmetry
breaking\cite{PhysRevB.83.184205,PhysRevB.93.094204}, which describes
the environment around each particle, offers an innovative attempt. In a
recent work\cite{jiang2026microscopicoriginbosonpeak}, a BP was found in
the ``simplest'' model: a scalar spring network with fluctuation of
coordination number on nodes, where the coordination number gives the
number of nodes connected to the central node. Local inversion symmetry
breaking is direct here due to the fluctuation of coordination number.
However, the mathematical mechanism by which the fluctuation of
coordination number leads to the salient features of BP formation has
not yet been established.

Inspired by
Refs.\cite{PhysRevB.93.094204,jiang2026microscopicoriginbosonpeak,Nie2017},
we analytically study the effect of defects that induce
coordination-number fluctuation in a two-dimensional square-lattice
scalar spring network. We find that deviations of the coordination
number \(C_N=3(5)\) from the perfect square lattice \(C_N=4\) result in
shifts of the dynamical response frequencies toward lower and higher
values, respectively. Those shifted toward zero finally form an
anomalous accumulation above zero, which is the BP. The shift and
accumulation of the dynamical response are guaranteed at the level of
mathematical principles by Weyl's theorem for Hermitian matrices: the
negative/positive semi-definiteness of the defect modification matrix
fixes the direction of the eigenvalue shift, while the high rank of the
multi-defect modification relaxes the interlacing constraint and allows
the eigenvalue distribution to be reshaped. Reference~\cite{jiang2026microscopicoriginbosonpeak}
demonstrates that coordination-number fluctuation serves as the
dominant feature leading to a BP. This work provides a theoretical
explanation that bridges the gap between the presence of defects and the
observable features of the VDOS, and thus contributes to our
understanding of vibrational anomalies in amorphous solids. The proposed
method and conclusions are general and can be extended to real-world
scenarios because they rest only on Weyl's theorem rather than on
perturbative or phenomenological assumptions.

\section{Theory and analysis}

Let us consider a scalar spring network implemented on a two-dimensional
square lattice. Each node is connected to its nearest neighbors by
springs of constant strength \(t\). The scalar dynamical matrix of the
perfect crystal network \(D_0\) is constructed as follows. The
off-diagonal elements are \(D_{0,ij}=-t\) if the \(i\)th and \(j\)th
nodes are connected by a spring, and the diagonal elements are
\(D_{0,ii}=-\sum_{j\neq i}D_{0,ij}\) by the requirement of mechanical
stability. The eigenvalues of \(D_0\) satisfy
\(D_0\,e_i=z_i\,e_i\), with \(z_i=\omega_i^2\), and the vibrational
density of states (VDOS) is
\(\rho(\omega)=\sum_i\delta(\omega_i-\omega)\). The indices
\(i,j\) run over all nodes from \(1\) to \(N\).

Disorder is introduced by displacing each node from its perfect
lattice site by a distance \(\delta<a\) along a random direction (\(a\)
is the lattice constant), then breaking connections whose length exceeds
\(a+\delta\) and forming new connections whose length falls below
\(a+\delta\)\cite{jiang2026microscopicoriginbosonpeak}. The dynamical
matrix of the resulting amorphous network is denoted by \(D\). This
construction is free of phenomenological input and preserves the
essential features of the VDOS, including Debye's law and BP at low
frequency and the Van Hove peak at high frequency, while reproducing the
coordination-number distribution observed in
simulations\cite{jiang2026microscopicoriginbosonpeak}. Since the springs
are scalar and of constant strength, the only remaining feature that
distinguishes the amorphous network from the crystal is the fluctuation
of coordination number. This motivates the central question of this
work: how does a local change of coordination number reshape the
low-frequency scaling of the VDOS?

\begin{figure}[htb]
    \centering
    \includegraphics[width=0.8\linewidth]{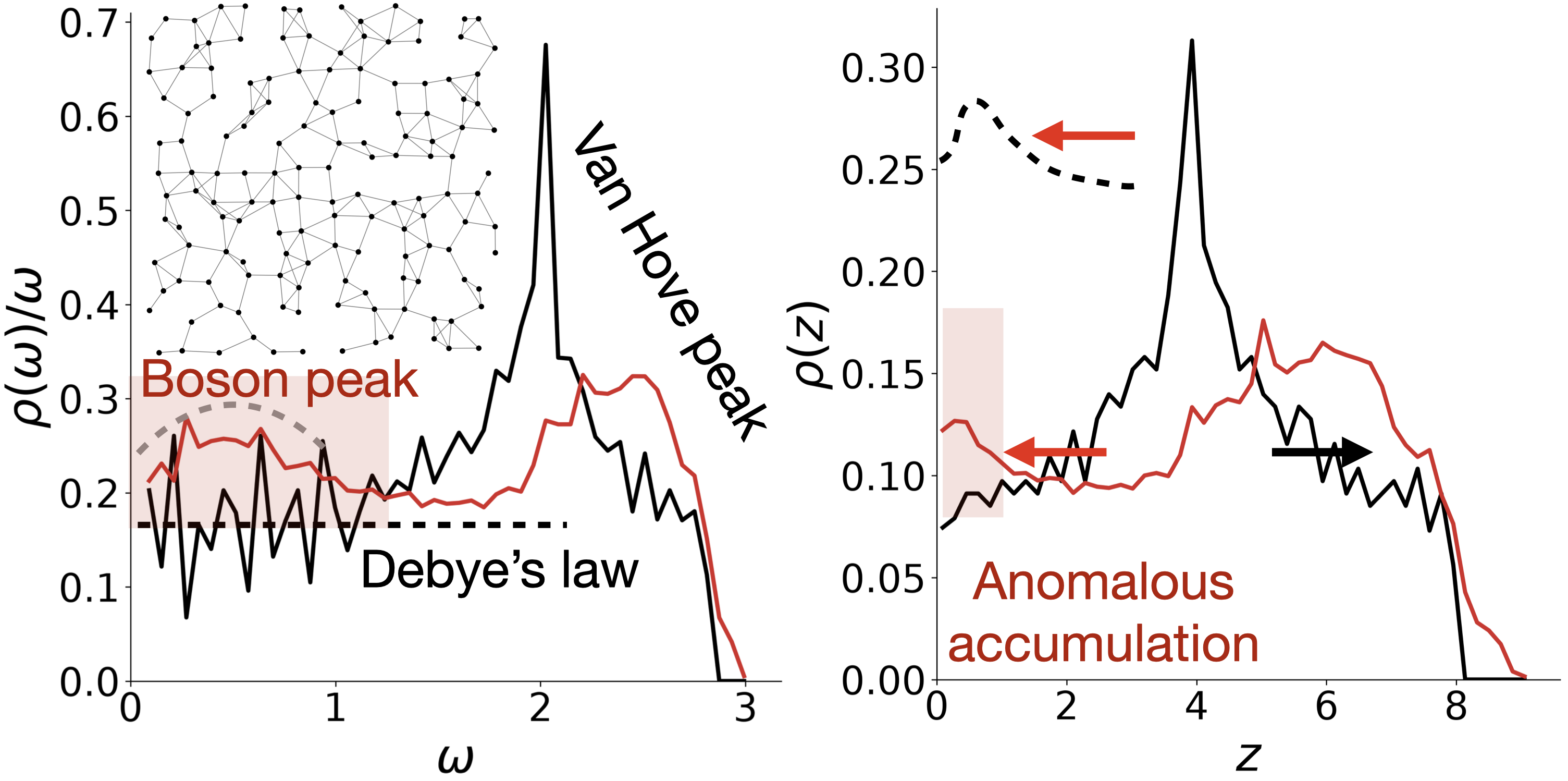}
    \caption{\textbf{Left}, reduced VDOS of a two-dimensional scalar
    spring network on a square lattice (black) and of the amorphous
    network (red) with the structure shown in the insert. The amorphous
    structure is generated as in Ref.\cite{jiang2026microscopicoriginbosonpeak}
    with parameters \(\sigma=0.3\) and \(t=1\). The dynamical features,
    including Debye's law, the Boson peak, and the Van Hove peak, are
    indicated accordingly. \textbf{Right}, distribution of eigenvalues
    of the dynamical matrices \(D_0\) for the square lattice (black) and
    \(D\) for the amorphous structure (red). The anomalous accumulation
    of small eigenvalues is highlighted in gray; it is the seed of the
    Boson peak after transforming into the frequency domain. The system
    size is \(60\times60\), and hence the total number of nodes is
    \(N=3600\). Each curve is averaged over \(5\) independent random
    seeds of the displacement directions. Periodic boundary conditions
    are applied throughout this work.}
    \label{fignewbp}
\end{figure}

To analyze the effect of coordination-number fluctuation, we write the
dynamical matrix of the amorphous network as
\begin{equation}
    D = D_0 + V,
\end{equation}
where \(V\) is the defect modification matrix. This decomposition is
exact. The matrix \(V\) collects all changes of the network connectivity:
removing a connection contributes a negative semi-definite term, while
adding a connection contributes a positive semi-definite term. In
general, \(V\) can be written as
\begin{equation}
    V = \sum_k \sigma_k\,|u_k\rangle\langle u_k|,
    \qquad \sigma_k = \mp t,
\end{equation}
where the negative sign corresponds to removing a connection and the
positive sign to adding one. For example, removing the connection between the \(6\)th and \(7\)th
nodes is represented by
\begin{equation}
    V = -t\bigl(|6\rangle-|7\rangle\bigr)
        \bigl(\langle6|-\langle7|\bigr).
\end{equation}
This matrix is negative semi-definite: setting
\(|u\rangle=|6\rangle-|7\rangle\), we have
\(V=-t\,|u\rangle\langle u|\). For any vector \(|\psi\rangle\),
\begin{equation}
    \langle\psi|V|\psi\rangle
    =
    -t\,\langle\psi|u\rangle\langle u|\psi\rangle
    =
    -t\,|\langle u|\psi\rangle|^2
    \le 0,
\end{equation}
since \(t>0\). Therefore \(V\preceq 0\) in the Loewner order\cite{matrix1997,matrix2002}. Similarly,
adding a connection gives \(V=+t\,|u\rangle\langle u|\succeq 0\).

The eigenvalues of \(D\) are denoted by \(z_i(D)\), ordered
increasingly, and similarly for \(D_0\). The central quantity is the
distribution of these eigenvalues, because the VDOS in the frequency
domain follows directly from \(z_i=\omega_i^2\). The effect of \(V\) is
analyzed using Weyl's theorem for Hermitian matrices, which provides
exact inequalities between the eigenvalues of \(D_0\) and those of
\(D_0+V\). Two properties of \(V\) are essential: its semi-definiteness
fixes the direction of the eigenvalue shift, and its rank controls how
many eigenvalues can be moved and whether their ordering can change.

We first consider the direction of the shift. If only connections are
removed, then every term in \(V\) is negative semi-definite, so
\(V\preceq 0\).
Hence
\begin{equation}
    D = D_0 + V \preceq D_0.
\end{equation}
By Weyl's monotonicity theorem\cite{matrix1997,matrix2002}, the
eigenvalues satisfy
\begin{equation}
    z_i(D) \le z_i(D_0), \qquad i=1,\dots,N, \label{eqshift}
\end{equation}
when both sets are ordered increasingly. Therefore all eigenvalues
shift toward smaller values or remain unchanged; none can shift toward
larger values. If only connections are added, then \(V\succeq 0\), so
\(D\succeq D_0\) and
\begin{equation}
    z_i(D) \ge z_i(D_0). \label{eqshiftup}
\end{equation}
Thus the direction of the spectral shift is fixed by the sign of the
defect modification matrix.

The zero mode is an exception in both cases. Because every row and
column of \(D_0\) and \(V\) sums to zero, the uniform translation mode
\(|1\rangle\) satisfies
\begin{equation}
    V|1\rangle = 0, \qquad D|1\rangle = 0,
\end{equation}
so the zero eigenvalue remains exactly at \(z=0\) regardless of the
defect.

Before analyzing how many eigenvalues can be moved, it is important to
take into account the degeneracy of the perfect-lattice spectrum. Let
\(\alpha\) label the distinct eigenvalues of \(D_0\), denoted by
\(z_\alpha(D_0)\), and let \(m_\alpha\) be the multiplicity of
\(z_\alpha(D_0)\), with \(\sum_\alpha m_\alpha=N\). The eigenvalues of
\(D_0\) are then \(z_\alpha(D_0)\) repeated \(m_\alpha\) times. In a
two-dimensional crystal, Debye's theory gives the low-frequency VDOS
\(\rho(\omega)\propto\omega\) for \(\omega\to0\). Since
\(z=\omega^2\), the corresponding eigenvalue density satisfies
\begin{equation}
    \rho(z) = \rho(\omega)\left|\frac{d\omega}{dz}\right|
    \propto \omega \cdot \frac{1}{2\omega}
    = \text{constant}.
\end{equation}
Thus, in the low-frequency region, the number of eigenvalues per unit
interval of \(z\) is approximately constant. In the thermodynamic
limit, the distinct eigenvalues \(z_\alpha(D_0)\) become dense, and
their multiplicities \(m_\alpha\) grow with the system size. This
degeneracy is not a peculiarity of any particular lattice but a direct
consequence of Debye's law in two dimensions.

We now examine how many eigenvalues can be moved and whether their
ordering can change. A single defect corresponds to removing or adding
one connection. The modification matrix is then rank one,
\begin{equation}
    V = \sigma\,|u\rangle\langle u|,
    \qquad \sigma = \mp t.
\end{equation}
For a rank-one Hermitian defect modification matrix, Weyl's interlacing theorem
applies\cite{matrix1997,matrix2002}. For a negative rank-one matrix \(V\) (removing a connection),
one has
\begin{equation}
    z_{i-1}(D_0) \le z_i(D) \le z_i(D_0),
    \label{eqinterlaceminus}
\end{equation}
with the convention \(z_0(D_0)=-\infty\). For a positive rank-one matrix \(V\) (adding a connection), one has
\begin{equation}
    z_i(D_0) \le z_i(D) \le z_{i+1}(D_0),
    \label{eqinterlaceplus}
\end{equation}
with \(z_{N+1}(D_0)=+\infty\). Here \(i\) labels all eigenvalues in
increasing order, including degenerate ones. These inequalities mean
that each eigenvalue can move only within the interval bounded by its
immediate neighbors in the perfect lattice. It cannot cross an adjacent
eigenvalue. In particular, if \(z_{i-1}(D_0)=z_i(D_0)=z_\alpha(D_0)\),
the inequality forces \(z_i(D)=z_\alpha(D_0)\), so that eigenvalue
cannot move at all.

This has a direct consequence for a degenerate subspace of multiplicity
\(m_\alpha\). Since \(V\) projected onto that subspace is a rank-one
matrix, a single defect can move at most one eigenvalue within the
subspace; the remaining \(m_\alpha-1\) eigenvalues stay at
\(z_\alpha(D_0)\). Therefore, a single defect cannot change the
ordering of the distinct eigenvalues \(z_\alpha(D_0)\) and cannot
reshape the distribution pattern of the spectrum. In Supplementary information, we provide independent analytical derivation of the above conclusions through Green's function method. To move all
\(m_\alpha\) eigenvalues of a degenerate subspace, the defect
modification matrix must have rank at least \(m_\alpha\).

When several connections are removed or added, the situation changes
qualitatively. The defect modification matrix becomes a sum of rank-one
terms,
\begin{equation}
    V = \sum_{k=1}^{r} \sigma_k\,|u_k\rangle\langle u_k|,
\end{equation}
where \(r=\operatorname{rank}(V)\) is the number of independent defect
directions. If the removed connections dominate, \(V\preceq 0\) and the
Weyl interlacing relation generalizes to
\begin{equation}
    z_{i-r}(D_0) \le z_i(D) \le z_i(D_0),
    \label{eqhighrankminus}
\end{equation}
with \(z_{i-r}(D_0)=-\infty\) for \(i-r\le 0\). If the added
connections dominate, \(V\succeq 0\) and
\begin{equation}
    z_i(D_0) \le z_i(D) \le z_{i+r}(D_0).
    \label{eqhighrankplus}
\end{equation}
Thus a rank-\(r\) perturbation allows the \(i\)th eigenvalue to move
across as many as \(r\) original eigenvalues. When \(r>1\), the
interlacing constraint is relaxed, and the ordering of the distinct
eigenvalues can change. In particular, if \(r\) is comparable to the
multiplicity \(m_\alpha\) of a low-frequency degenerate subspace, the
entire subspace can be shifted, and eigenvalues originally located at
higher values can be moved below lower ones. The spectral distribution
is thereby reshaped.

This is the central mathematical mechanism. A single defect has
\(r=1\) and is therefore locked into the interval between neighboring
perfect-lattice eigenvalues; it can move at most one state within each
degenerate subspace. Multiple defects make \(V\) a high-rank matrix,
\(r>1\), and thereby provide the necessary freedom for the eigenvalue
distribution to be reorganized. In the thermodynamic limit, the
distinct eigenvalues in the low-frequency region become dense, and
their multiplicities become large. To move an entire degenerate
subspace of multiplicity \(m_\alpha\), one needs at least
\(r\sim m_\alpha\) independent defect directions. Hence a sufficiently
high density of defects is required to reshape the low-energy part of
the spectrum. The resulting accumulation of eigenvalues just above
\(z=0\), as shown in the right panel of Fig.\ref{fignewbp}, is the
theoretical origin of the anomalous low-frequency VDOS, that is, the BP.

The above analysis is general. It does not rely on the specific size of
the lattice, the specific positions of the defects, or any perturbative
expansion. Three ingredients enter the framework: the semi-definiteness
of the defect modification matrix, which fixes the direction of the
eigenvalue shift; the degeneracy of the perfect-lattice spectrum, which
provides a dense set of distinct eigenvalues with large multiplicities
in the thermodynamic limit; and the high rank of the multi-defect
modification, which relaxes the interlacing constraint and allows the
eigenvalue distribution to be reshaped. All these ingredients are
analyzed using Weyl's theorem for Hermitian matrices, which provides
exact and non-perturbative inequalities. The first two ingredients
follow from the structure of the scalar spring network and Debye's law
in two dimensions, while the third emerges whenever enough defects are
present to make \(V\) a high-rank matrix. The framework is therefore
applicable to real amorphous solids, where the fluctuation of
coordination number is the dominant microscopic feature\cite{jiang2026microscopicoriginbosonpeak}, and provides a
route from local coordination defects to the anomalous VDOS that rests
only on mathematical principles.

\section{Discussion and outlook}
The analysis above leads to two natural corollaries. The first is that
the BP is a collective dynamical phenomenon involving the majority of
particles, rather than the effect of a few isolated defects. A single
defect, or a small number of well-separated defects, cannot reshape the
distribution of eigenvalues: each defect has rank one and can move at
most one eigenvalue within a degenerate subspace, while the ordering of
the distinct eigenvalues remains fixed. Only when defects are dense
enough to make the modification matrix high-rank does the eigenvalue
distribution become reorganized beyond degenerate subspaces. This is consistent with previous
reports that the BP emerges from the cooperative dynamics of many local
environments rather than from a few localized objects\cite{qx17-q53j}.

The second corollary is that there exists a minimum threshold for the
number of defects required to produce a BP. In two dimensions, this
threshold is approximately
\(N_{\text{defect}}^{\min}\sim \rho(z\to 0)N\), where
\(\rho(z\to 0)\) is the low-eigenvalue density of the perfect lattice
and \(N\) is the total number of particles. The estimate follows from
the rank-one nature of a single defect: one defect can move only one
state within a degenerate subspace. Therefore, \(\rho(z\to 0)N\)
defects are needed to move all states in the low-eigenvalue region.
According to Fig.\ref{fignewbp}, \(\rho(z\to 0)\approx 0.1\), which
gives a defect threshold consistent with the order of magnitude reported
in Ref.\cite{Nie2017}.

The mechanism by which coordination-number fluctuations reshape the VDOS
can now be summarized in three ingredients. First,
Eqs.\eqref{eqshift} and \eqref{eqshiftup}: the defect modification
matrix is positive or negative semi-definite; in particular, when the
coordination number is lower than the average, the removed connections
give a negative semi-definite \(V\) that shifts the eigenvalues
downward. Second, Eqs.\eqref{eqhighrankminus} and
\eqref{eqhighrankplus}: the high rank of the multi-defect modification
relaxes the interlacing constraint, so that eigenvalues can cross over
one another and accumulate at low frequency. Third, the positive
semi-definiteness of the full dynamical matrix \(D\) guarantees that the
zero eigenvalue remains at zero and that no eigenvalue can shift below
zero. These three elements together provide the complete set of
conditions for the accumulation of eigenvalues above \(z=0\), which is
the origin of the BP.

In the present analysis, defects are classified as either removed or
added connections, and the two cases are shown to shift the eigenvalues
downward and upward, respectively. In real amorphous solids, however,
both types of defects coexist, as illustrated in the insert of the left
panel of Fig.\ref{fignewbp} and discussed in
Ref.\cite{jiang2026microscopicoriginbosonpeak}. From the observed
distribution of eigenvalues, the impacts of the two types of defects on
the direction of the eigenvalue shift are approximately independent. As
shown in the right panel of Fig.\ref{fignewbp}, the anomalous
accumulations of \(\rho(z)\) in the amorphous structure relative to the
crystal structure appear on both the left and right sides of the \(z\)
range. The left accumulation is the seed of the BP, while the right one
lies beyond the original Van Hove peak.

The opposite directions in which removing and adding connections reshape
the eigenvalue distribution lead to a further corollary. Pure removal of
connections produces an anomalous accumulation of eigenvalues near zero
and hence a BP in the frequency domain, as confirmed in
Ref.\cite{Nie2017}. Pure addition of connections, by contrast, shifts
eigenvalues toward larger values and therefore produces no BP. This
corollary is verified at the end of the Supplementary Information.
Adding connections between nodes is itself a way of creating disorder.
The absence of a BP in such a disordered solid raises a valuable open
question for real-world disordered systems.

The two-dimensional scalar spring network is a highly simplified model,
retaining essentially only the fluctuation of coordination number as the
relevant microscopic feature. Nevertheless, the underlying analysis of
the eigenvalues of the dynamical matrix is of considerable generality in
the study of amorphous-solid dynamics, ranging from zero-temperature
dynamical simulations\cite{Xi2026,Hu2022} to experimentally observed
active granular matter\cite{PhysRevLett.133.188302}. Other factors, such
as temperature, elastic inhomogeneity, and different defect types, may
also influence the BP and are natural directions for further study. At
zero temperature, however, amorphous-solid models have already been
widely used as a platform for studying the BP \cite{Xi2026,Hu2022}.
Elastic inhomogeneity has been argued not to be a necessary ingredient
\cite{Xi2026,jiang2026microscopicoriginbosonpeak} and mainly introduces
damping. Moreover, different defect types, and even the structures of
real amorphous solids obtained from simulation, can be approximated by
the fluctuation of node coordination number in a spring network
\cite{Nie2017,jiang2026microscopicoriginbosonpeak}. These results
suggest that coordination-number fluctuation is likely the most
fundamental mechanism underlying the BP after all unnecessary details
are stripped away. The present work directly clarifies, at the level of
mathematical principles, the mechanism by which coordination-number
fluctuation reshapes the VDOS into a BP.


\clearpage

\setcounter{equation}{0}
\setcounter{figure}{0}
\setcounter{table}{0}
\renewcommand{\theequation}{S\arabic{equation}}
\renewcommand{\thefigure}{S\arabic{figure}}
\renewcommand{\thetable}{S\arabic{table}}

\section*{\Large Supplementary Information}
\noindent\textbf{\large Weyl's theorem and the theoretical origin of Boson peak in two-dimensional scalar spring network}\\[0.5em]
Cunyuan Jiang\\
School of Physics, Zhengzhou University, Zhengzhou 450001, Henan, China\\[0.5em]
\noindent\textbf{Abstract:} In this supplementary material, we provide full derivation details toward the conclusion of how defects shift the eigen frequency of dynamical response.

\medskip

For a two-dimensional crystalline scalar spring network, the dynamical matrix can be constructed according to the connections between nodes. The off-diagonal elements are \(D_{0,ij}=-t\) if the \(i\)-th and \(j\)-th nodes are connected (nearest neighbors) by a spring with strength \(t\). The diagonal elements are \(D_{0,ii}=-\sum_{j\neq i}D_{0,ij}\) by the requirement of stability.

Define the Green's function as,
\begin{equation}
G_0(z)=(zI-D_0)^{-1},\qquad z=\omega^2-i0^+.
\end{equation}

Let
\begin{equation}
D=D_0+V,\qquad
G(z)=(zI-D_0-V)^{-1}.
\end{equation}
Here \(V\) denotes the disorder modification of the connections in the crystalline structure. From the Dyson equation \(G^{-1}=G_0^{-1}-V\), multiplying by \(G_0\) on the left and by \(G\) on the right gives
\begin{equation}
    G_0 G^{-1} G = G_0 G_0^{-1} G - G_0 V G.
\end{equation}
Using \(G^{-1}G=I\) and \(G_0 G_0^{-1}=I\), this gives
\begin{equation}
    G = G_0 + G_0 V G.
\end{equation}
Equivalently,
\begin{equation}
    (I-G_0V)G=G_0,
\end{equation}
so that
\begin{equation}
    G=(I-G_0V)^{-1}G_0=G_0(I-VG_0)^{-1}.
\end{equation}

Let \(P\) project onto the defect subspace and let
\begin{equation}
    P^2=P,\qquad P^\dagger=P,\qquad Q=I-P,
\end{equation}
so that
\begin{equation}
    Q^2=Q,\qquad PQ=QP=0.
\end{equation}
Since \(V\) modifies only the nodes and connections that involve defects, we have
\begin{equation}
    V=PVP.
\end{equation}
Here \(Q\) projects onto the remaining nodes without defects. Then
\begin{equation}
    QV=QPVP=0,\qquad VQ=PVPQ=0.
\end{equation}
We now evaluate \(I-G_0V\) in the \(P\oplus Q\) basis. First decompose \(G_0V\) as
\begin{equation}
    G_0V=(P+Q)G_0V(P+Q).
\end{equation}
Expanding,
\begin{equation}
    G_0V
    =
    P G_0 V P
    + P G_0 V Q
    + Q G_0 V P
    + Q G_0 V Q.
\end{equation}
Since \(VQ=0\), the terms containing \(G_0VQ\) vanish:
\begin{equation}
    P G_0 V Q=0,\qquad Q G_0 V Q=0.
\end{equation}
Therefore
\begin{equation}
    G_0V=P G_0 V P+Q G_0 V P.
\end{equation}
Hence
\begin{equation}
    I-G_0V
    =
    I-P G_0 V P-Q G_0 V P.
\end{equation}
In the \(P\oplus Q\) basis, \(I\) is block diagonal:
\begin{equation}
    I=
    \begin{pmatrix}
        I_P & 0\\
        0 & I_Q
    \end{pmatrix},
\end{equation}
and \(P G_0 V P\) and \(Q G_0 V P\) act only in the indicated blocks:
\begin{equation}
    P G_0 V P=
    \begin{pmatrix}
        P G_0 V P & 0\\
        0 & 0
    \end{pmatrix},
    \qquad
    Q G_0 V P=
    \begin{pmatrix}
        0 & 0\\
        Q G_0 V P & 0
    \end{pmatrix}.
\end{equation}
Thus,
\begin{equation}
    I-G_0V=
    \begin{pmatrix}
        I_P-P G_0 V P & 0\\
        -Q G_0 V P & I_Q
    \end{pmatrix}.
\end{equation}
This is a block lower-triangular matrix of the form
\begin{equation}
    \begin{pmatrix}
        A & 0\\
        C & D
    \end{pmatrix},
    \qquad
    A=I_P-P G_0 V P,\quad
    C=-Q G_0 V P,\quad
    D=I_Q.
\end{equation}
Its inverse is
\begin{equation}
    \begin{pmatrix}
        A & 0\\
        C & D
    \end{pmatrix}^{-1}
    =
    \begin{pmatrix}
        A^{-1} & 0\\
        -D^{-1} C A^{-1} & D^{-1}
    \end{pmatrix}.
\end{equation}
Since \(D=I_Q\), we have \(D^{-1}=I_Q\), and therefore
\begin{equation}
    (I-G_0V)^{-1}
    =
    \begin{pmatrix}
        (I_P-P G_0 V P)^{-1} & 0\\
        Q G_0 V P (I_P-P G_0 V P)^{-1} & I_Q
    \end{pmatrix}.
\end{equation}
In operator form, this is
\begin{equation}
    (I-G_0V)^{-1}
    =
    P(I-P G_0 V P)^{-1}P
    +Q G_0 V P (I-P G_0 V P)^{-1}P
    +Q.
\end{equation}
Now use
\begin{equation}
    P+Q G_0 V P
    =
    P+(I-P)G_0 V P
    =
    P+G_0 V P-P G_0 V P.
\end{equation}
Since
\begin{equation}
    G_0 V P
    =
    (P G_0 V P+Q G_0 V P)P
    =
    P G_0 V P+Q G_0 V P,
\end{equation}
we obtain
\begin{equation}
    P+Q G_0 V P
    =
    P+G_0 V P-P G_0 V P
    =
    P+Q G_0 V P.
\end{equation}
This allows the compact rewriting
\begin{equation}
    (I-G_0V)^{-1}
    =
    I+G_0V(I-P G_0 V P)^{-1}P.
\end{equation}
Finally, substituting this into \(G=(I-G_0V)^{-1}G_0\) gives
\begin{equation}
    G
    =
    \left[
    I+G_0V(I-P G_0 V P)^{-1}P
    \right]G_0.
\end{equation}
Hence
\begin{equation}
    G
    =
    G_0
    +
    G_0 V (I-P G_0 V P)^{-1} P G_0.
\end{equation}
The poles related to defect are therefore determined by,
\begin{equation}
    \det(I-P G_0 V P)=0.
\end{equation}
The projection matrix \(P\) reduces the determinant to defect subspace, significantly simplifying the original condition \(\det(I-G_0V)=0\), which involves all nodes.

\vspace{2em}
\noindent\textbf{Removing a bond between the nearest neighbors}

Let us consider a \(4\times4\) square lattice with periodic boundary conditions. The nodes are labeled \(1,2,3,4\) for the first row, \(5,6,7,8\) for the second row, and so on. For the first situation, we remove the nearest-neighbor connection between the \(6\)th and \(7\)th nodes. The disorder modification \(V\) is therefore
\begin{equation}
V
=
t\bigl(|6\rangle\langle7|+|7\rangle\langle6|\bigr)
-
t\bigl(|6\rangle\langle6|+|7\rangle\langle7|\bigr).
\end{equation}
Here \(|i\rangle\) is the \(i\)th basis vector of the node space. The projection matrix onto the defect subspace is
\begin{equation}
P=|6\rangle\langle6|+|7\rangle\langle7|.
\end{equation}
Because of symmetry of the Green's function matrix, we have
\begin{equation}
G_{0,66}=G_{0,77},\qquad
G_{0,67}=G_{0,76}.
\end{equation}
Therefore
\begin{equation}
PG_0VP
=
t
\begin{pmatrix}
\Delta & -\Delta\\
-\Delta & \Delta
\end{pmatrix},
\qquad
\Delta=G_{0,67}-G_{0,66}.
\end{equation}
Thus,
\begin{equation}
I-PG_0VP
=
\begin{pmatrix}
1-t\Delta & t\Delta\\
t\Delta & 1-t\Delta
\end{pmatrix},
\end{equation}
and
\begin{equation}
\det (I-PG_0VP)=(1-t\Delta)^2-(t\Delta)^2=1-2t\Delta.
\end{equation}
The pole condition is
\begin{equation}
G_{0,67}-G_{0,66} =\frac{1}{2t}.
\end{equation}

For the perfect \(4\times4\) periodic square lattice with \(t=1\), the eigenvalues are
\begin{equation}
    \lambda_{k_x,k_y}
    =
    4-2(\cos k_x+\cos k_y),
    \qquad
    k_x,k_y\in\left\{0,\frac{\pi}{2},\pi,\frac{3\pi}{2}\right\}.
\end{equation}
Define
\begin{equation}
    a(k_x)=2-2\cos k_x,\qquad
    b(k_y)=2-2\cos k_y.
\end{equation}
Then
\begin{equation}
    \lambda=a+b,
    \qquad
    a,b\in\{0,2,4\},
\end{equation}
with multiplicities
\begin{equation}
    m(0)=1,\qquad m(2)=2,\qquad m(4)=1.
\end{equation}
Therefore the perfect square lattice on-site Green's function has the partial-fraction representation
\begin{equation}
    G_{0,66} (z)
    =
    \frac{1}{16}
    \left(
    \frac{1}{z}
    +\frac{4}{z-2}
    +\frac{6}{z-4}
    +\frac{4}{z-6}
    +\frac{1}{z-8}
    \right).
\end{equation}

Because nodes \(6\) and \(7\) are horizontally adjacent, so that \(\Delta\mathbf{r}=\mathbf{r}_6-\mathbf{r}_7=(1,0)\), we have
\begin{equation}
   G_{0,67} (z)
    =
    \frac{1}{16}\sum_{k_x,k_y}
    \frac{e^{ik_x}}{z-\lambda_{k_x,k_y}}.
\end{equation}
Summing over \(k_y\) and then over \(k_x\), the terms with \(e^{ik_x}=\pm i\) cancel, and one obtains
\begin{equation}
    G_{0,67} (z)
    =
    \frac{1}{16}
    \left(
    \frac{1}{z}
    +\frac{2}{z-2}
    -\frac{2}{z-6}
    -\frac{1}{z-8}
    \right).
\end{equation}
Hence,
\begin{equation}
\begin{aligned}
    G_{0,67}-G_{0,66}
    &=
    \frac{1}{16}
    \left(
    -\frac{2}{z-2}
    -\frac{6}{z-4}
    -\frac{6}{z-6}
    -\frac{2}{z-8}
    \right) \\
    &=
    -\frac{1}{8}
    \left(
    \frac{1}{z-2}
    +\frac{3}{z-4}
    +\frac{3}{z-6}
    +\frac{1}{z-8}
    \right).
\end{aligned}
\end{equation}
The pole condition
\begin{equation}
    G_{0,67}-G_{0,66}=\frac{1}{2}
\end{equation}
therefore becomes
\begin{equation}
    \frac{1}{z-2}
    +\frac{3}{z-4}
    +\frac{3}{z-6}
    +\frac{1}{z-8}
    =
    -4.
\end{equation}
The above pole equation has \(4\) solutions near the original poles.

Let
\begin{equation}
    F(z) = \frac{1}{z-2}
    +\frac{3}{z-4}
    +\frac{3}{z-6}
    +\frac{1}{z-8},
\end{equation}
its first-order derivative with respect to \(z\) is
\begin{equation}
    F(z)' = - \frac{1}{(z-2)^2}
    -\frac{3}{(z-4)^2}
    -\frac{3}{(z-6)^2}
    -\frac{1}{(z-8)^2} < 0.
\end{equation}
Because the first-order derivative is negative, \(F(z)\) is monotonically decreasing on each interval where it is defined. Note that \(F(z)\) is not defined at the poles \(z=2,4,6,8\). In the domain \(z\in(-\infty,2)\), the function \(F(z)\) monotonically decreases from \(F(-\infty)=0^-\) to \(F(2^-)=-\infty\), and therefore must have a unique solution \(z_2<2\) satisfying \(F(z_2)=-4\). The position of \(z_2\) can be estimated by solving
\begin{equation}
        \frac{1}{z_2 -2}
    +\frac{3}{2-4}
    +\frac{3}{2-6}
    +\frac{1}{2-8}
    =
    -4.
\end{equation}
Rewriting the above equation as follows, \(z_2\) can be obtained directly as
\begin{equation}
    z_2 \approx - \left(\dfrac{3}{2-4} + \dfrac{3}{2-6} + \dfrac{1}{2-8} + 4\right)^{-1} + 2 = 1.37.
\end{equation}
One of the exact solutions of \(F(z)=-4\) is \(1.50\), which approximately agrees with the estimation.

The other solutions of \(F(z)=-4\) lie in the domains \(z\in(2,4)\), \(z\in(4,6)\), and \(z\in(6,8)\), and can be estimated similarly. In general, the solutions of the pole equation \(F(z)=-4\) have the following form
\begin{equation}
    z_i \sim z_i^{(0)} + \alpha m ,\qquad \alpha <0,
\end{equation}
where \(z_i^{(0)}\) denotes the poles of the lattice on-site Green's function \(G_{0,66}\), and \(m\) denotes the multiplicity of the poles \(z_i^{(0)}\). \(\alpha\) is the shift of the new poles induced by removing a connection from the perfect lattice relative to the original poles. In the case of removing a connection, the shift \(\alpha\) is always negative due to the monotonicity of \(F(z)\) and the negative sign on the right-hand side of the pole equation \(F(z)=-4\). Therefore, the effect of removing a connection from the perfect lattice can be summarized as shifting some poles of the perfect lattice toward smaller values.

The multiplicity of the poles of the perfect lattice is usually greater than \(1\). For example, according to Debye theory, the vibrational density of states of an infinite crystal should be \(\rho(\omega)\propto\omega\) in the low-frequency region for the two-dimensional case. Since \(z=\omega^2-i0^+\), we have \(\rho(\lambda)=\rho(\omega)\left|d\omega/d\lambda\right|\), and \(\rho(\lambda)\propto\mathrm{const.}\) for small \(\lambda\). Then a question must be answered. As we have concluded that some poles of the perfect lattice will be shifted toward smaller values due to breaking a connection between two nodes, how many of the multiple poles will be shifted?

The defect modification matrix can be written as a single outer product,
\begin{equation}
    V
    =
    -t\bigl(|6\rangle-|7\rangle\bigr)
    \bigl(\langle6|-\langle7|\bigr),
\end{equation}
Hence \(V\) has rank \(1\). Let \(P_n\) be the projector onto a degenerate subspace of \(D_0\). The first-order perturbation of defect is \(M_n=P_nVP_n\). Since \(\operatorname{rank}(M_n)\le\operatorname{rank}(V)=1\), \(M_n\) has at most one nonzero eigenvalue. Therefore, in each degenerate subspace, at most one state receives a nonzero shift, while the remaining \(m-1\) states remain at their original positions.

The pole at zero remains exactly at zero due to the stability condition \(\sum_j D_{ij}=0\). More strictly, no pole is allowed to shift below zero, even though a substantial part of the poles will be shifted toward smaller values when a connection is removed from the perfect lattice.

\vspace{2em}
\noindent\textbf{Adding a bond between the next-nearest neighbors}

Similarly, a connection can be added between next-nearest neighbors, for example, between the \(6\)th and \(11\)th nodes. In this case, the disorder modification is
\begin{equation}
V
=
-t\bigl(|6\rangle\langle11|+|11\rangle\langle6|\bigr)
+
t\bigl(|6\rangle\langle6|+|11\rangle\langle11|\bigr).
\end{equation}
The projection is changed accordingly to
\begin{equation}
P=|6\rangle\langle6|+|11\rangle\langle11|,
\end{equation}
and the equality relation between elements of the lattice Green's function is
\begin{equation}
G_{0,66}=G_{0,11,11},\qquad
G_{0,6,11}=G_{0,11,6}.
\end{equation}

After an analysis similar to the previous case of removing a bond, the pole condition is determined as
\begin{equation}
G_{0,6,11}-G_{0,66}=-\frac{1}{2t}.
\end{equation}
By expressing the Green's functions in partial-fraction form,
\begin{equation}
    G_{0,66}
    =
    \frac{1}{16}
    \left(
    \frac{1}{z}
    +\frac{4}{z-2}
    +\frac{6}{z-4}
    +\frac{4}{z-6}
    +\frac{1}{z-8}
    \right),
\end{equation}
and
\begin{equation}
    G_{0,6,11}
    =
    \frac{1}{16}
    \left(
    \frac{1}{z}
    -\frac{2}{z-4}
    +\frac{1}{z-8}
    \right),
\end{equation}
the pole condition becomes
\begin{equation}
    \frac{1}{z-2}
    +\frac{2}{z-4}
    +\frac{1}{z-6}
    =
    2.
\end{equation}

Let
\begin{equation}
    F(z) = \frac{1}{z-2}
    +\frac{2}{z-4}
    +\frac{1}{z-6},
\end{equation}
its first derivative with respect to \(z\) is still negative:
\begin{equation}
    F(z)' = - \frac{1}{(z-2)^2}
    -\frac{2}{(z-4)^2}
    -\frac{1}{(z-6)^2} < 0.
\end{equation}
\(F(z)\) is still monotonically decreasing in the case of adding a connection, but note that the right-hand side of the pole condition \(F(z)=2\) is positive, whereas it was negative in the case of removing a connection. We estimate one of the solutions of the pole condition,
\begin{equation}
    \frac{1}{z_2 -2}
    +\frac{2}{2-4}
    +\frac{1}{2-6}
    =
    2,
\end{equation}
and
\begin{equation}
    z_2 \approx \left(2 - \dfrac{2}{2-4} - \dfrac{1}{2-6} \right)^{-1} + 2 = 2.31.
\end{equation}
Because of the monotonicity of \(F(z)\) and the positive sign on the right-hand side of \(F(z)=2\), the general form of the poles can be concluded as
\begin{equation}
    z_i \sim z_i^{(0)} + \alpha m ,\qquad \alpha >0,
\end{equation}
which means some poles will be shifted toward larger values due to the addition of a connection. The discussions about multiple poles and the zero pole are similar to the case of removing a connection. The zero pole will not be affected by defects due to the stability of the total system.

\vspace{2em}
\noindent\textbf{Effects of removing and adding connections}

\begin{figure}[htb]
    \centering
    \includegraphics[width=0.8\linewidth]{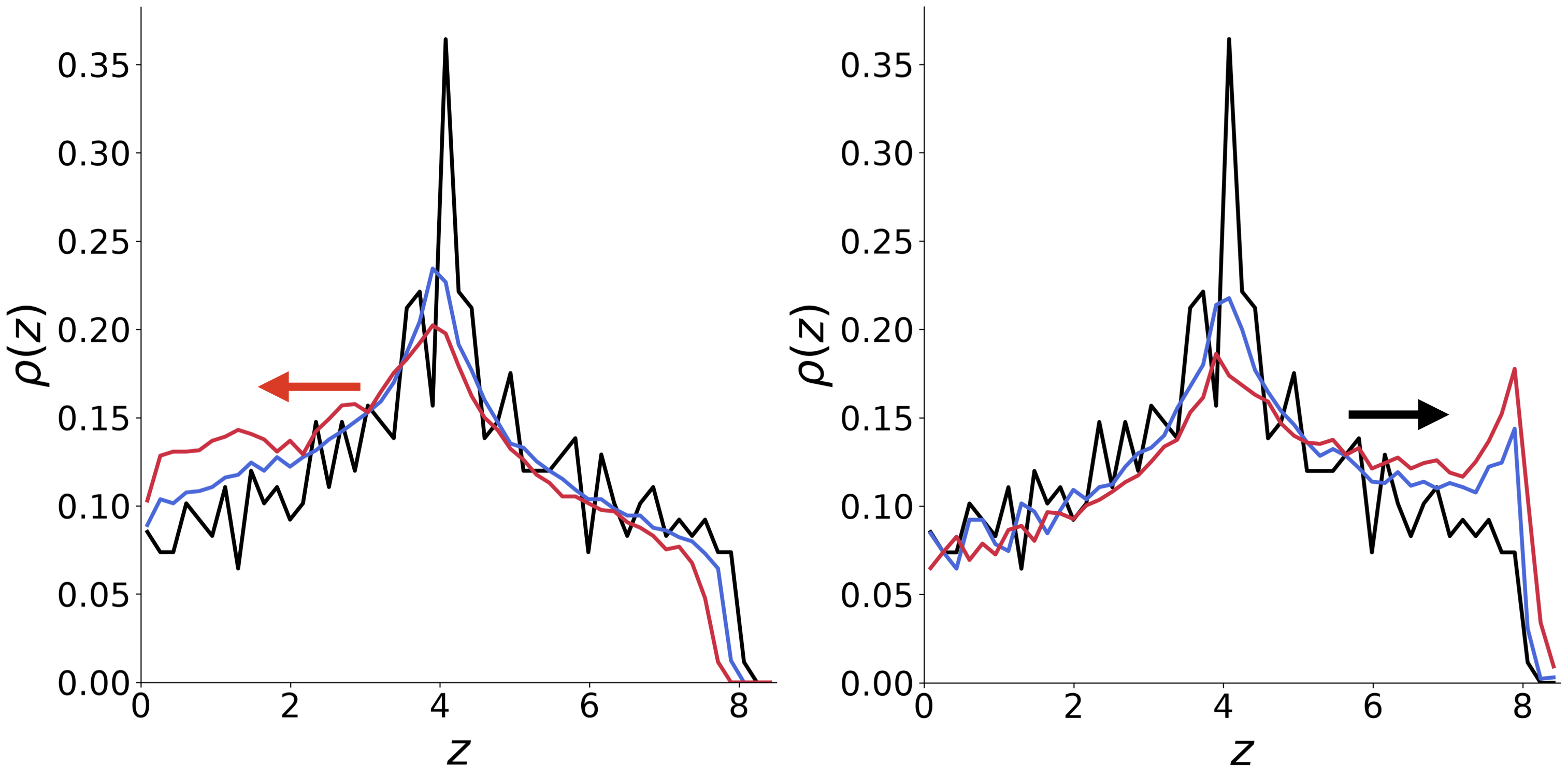}
    \caption{\textbf{Left}, eigenvalue distribution of a two-dimensional
    scalar spring network on a square lattice (black), and after randomly
    removing \(300\) (blue) and \(600\) (red) nearest-neighbor
    connections. The arrow points toward zero, indicating the anomalous
    accumulation of low eigenvalues that seeds the Boson peak.
    \textbf{Right}, eigenvalue distribution of the same perfect lattice
    (black), and after randomly adding \(300\) (blue) and \(600\) (red)
    next-nearest-neighbor connections. The arrow points toward larger
    eigenvalues, indicating an anomalous accumulation at high eigenvalues
    without a Boson peak.}
    \label{fig:addremove}
\end{figure}

The numerical results in Fig.\ref{fig:addremove} directly confirm the
theoretical prediction of the main text. Removing connections shifts the
poles downward and produces an anomalous accumulation of eigenvalues
near \(z=0\), which is the seed of the Boson peak after transforming to
the frequency domain. Adding connections, by contrast, shifts the poles
upward and produces an anomalous accumulation at large eigenvalues,
leaving the low-frequency region depleted rather than enhanced. This
asymmetry between removal and addition of connections is a direct
consequence of the semi-definiteness of the defect modification matrix
and the monotonicity of the eigenvalues under the Loewner order. It
explains why a Boson peak emerges in disordered solids where the
coordination number is reduced, while a purely connectivity-enhanced
disordered network does not exhibit a Boson peak.

\end{document}